# THE OBSERVATORY

## GROPING TOWARD LINEAR REGRESSION ANALYSIS: NEWTON'S ANALYSIS OF HIPPARCHUS' EQUINOX OBSERVATIONS

*By Ari Belenkiy*
Department of Statistics, Simon Fraser University, Vancouver BC, Canada
and
*Eduardo Vila Echagüe*
IBM, Santiago, Chile

　　In 1700 February, Isaac Newton needed a precise tropical year to design a new universal calendar to supercede the Gregorian one. However, 17th-Century astronomers were uncertain of the long-term variation in the inclination of the Earth's axis and were suspicious of Ptolemy's equinox observations. As a result, they produced a wide range of tropical years. This uncertainty led Newton to choose the ten equinox observations of Hipparchus of Rhodes as the most reliable among those available. Averaging the autumnal and vernal sets separately, he combined the results with Flamsteed's corresponding equinox observations, joining each pair with a line whose slope gave a deficiency of the tropical year *versus* the Julian year. After averaging the two, he corrected Flamsteed's year. Though Newton had a very limited sample of data, he obtained a tropical year only a few seconds longer than the average one between his and Hipparchus' time. As a by-product, Newton spotted, alongside Flamsteed, an error in the position of Hipparchus' equatorial ring, which was a matter of concern to later science.

　　Newton wrote down the first of the two so-called 'normal equations' known from the ordinary least-squares (OLS) method. In that procedure, Newton seems to have been the first to employ the *mean* (average) value of the data-set, while the other leading astronomers of the era (Tycho Brahe, Galileo, and Kepler) used the *median*. Fifty years after Newton, in 1750, Newton's method was rediscovered and enhanced by Tobias Mayer. Remarkably, the same regression method served with distinction in the 1920s when the founding fathers of modern cosmology, Georges Lemaître (1927), Edwin Hubble (1929), and Willem de Sitter (1930), employed it to derive the Hubble constant.

*Introduction: the dawn of regression analysis*

"The only thing which is surprising is that this principle [of the Least Squares], which suggests itself so readily that no particular value at all can be placed on the idea alone, was not already applied 50 or 100 years earlier by others, *e.g.*, Euler or Lambert or Halley or Tobias Mayer, although it may very easily be that the latter, for example, has applied that sort of thing without announcing it, just as every calculator necessarily invents a collection of devices and methods which he propagates by word of mouth only as occasion offers …" Gauss to Olbers, Göttingen, 1812 January 24[1].

The OLS (ordinary least-squares) regression is an optimization procedure that consists of taking several derivatives of a certain quantity and setting them equal to zero to get a set of linear ('normal') equations. However, until 1805, this procedure was not known and optimization was carried out in purely intuitive ways. In the 1748 prize-winning 123-page-long memoir *Recherches sur les irrégularités du mouvement de Saturne et de Jupiter*, published in Paris in 1749, Leonhard Euler, then the head of the Berlin Academy, arrived at 75 equations with eight uknowns but only half-heartedly proceeded combining observations to form a smaller set of equations, erroneously believing that the error "would multiply"[2]. In contrast, a year later, in 1750, the German astronomer Tobias Mayer, then a cartographer at the Homann Company in Nürnberg, studied the libration of the Moon over a period of one year, performing 27 observations of the crater Manilius, and obtained a system of 27 linear equations with three unknowns[3]. Splitting all the equations into three equal groups with similar characteristics and summing coefficients within each group, he arrived at a set of three linear equations, which he further solved in a standard 'Gaussian' way. Mayer's optimization procedure resulted in a system of three equations with dominant coefficients on the major diagonal, where, in Mayer's words, "the differences between the three sums are made as large as possible." The method later became known in Europe as Mayer's method or the 'method of averages'[4].

Averaging lies at the heart of the analytic part of the linear-regression method, though it is not so explicit in the modern least-squares technique. Remarkably, Mayer did not stop there but proceeded with a kind of error analysis, estimating that the combined error decreases in proportion to the number of combined equations[5]. Thus, Mayer's 1750 paper *Abhandlung über die Umwälzung des Monds um seine Axe und die scheinbare Bewegung der Mondsflecken* (Treatise on the rotation of the Moon on its axis and the apparent motion of the Moon spots) became a precursor for what later became known as regression analysis.

However, it is noteworthy that fifty years earlier than Mayer, in 1700, Isaac Newton had carried out similar averaging. Mayer's purely algebraic averaging can be viewed geometrically as finding the centre of gravity for three separate groups of points and then drawing a plane over them. For his part, Newton kept the geometrical picture from the very beginning. After separating two qualitatively distinct sets of points, autumnal and vernal equinoxes, he formed two regression lines that passed through the centre of gravity of each group and an *outlier*, later averaging their two slopes to form a single estimate. Though Newton did not come up with anything similar to an error analysis, there are signs that he felt he reduced the error by splitting the entire set into two groups.

Newton never published his method; it remained in a group of drafts known as *Yahuda 24*, now in the Jewish National Library in Jerusalem, first described by Belenkiy & Vila Echagüe in 2005[5]. In three of them, following a request from the Royal Society in 1700 February to respond to a letter by G. W. Leibniz, Newton constructed his civil (solar) and ecclesiastical (lunar) calendars. As a benchmark for the solar calendar, he needed to know the tropical year. In three successive drafts he chose three different values of $365^d\ 5^h\ 48^m$ and (*a*) $45^s$, (*b*) $48^s$, and (*c*) $56^s$ or $57^s$.

This paradoxical situation — the changing opinion about one of the *fundamental* astronomical 'constants' so quickly — demands an explanation. The simplest explanation is that at the end of the 17th Century the tropical year was uncertain within a wide range. Therefore, Section 1 describes the state of European astronomy by 1700. Section 2 explores what motivated Newton to choose the first two values. Section 3 presents the essence of the 'ordinary least-squares' regression and the two 'normal' equations needed to estimate the slope and intercept of the regression line. Section 4 examines Newton's method based on the first 'normal' equation that was responsible for the slope of the regression line. Section 5 addresses Newton's astronomical worldview in 1700. The afterword relates to how the same method served in modern cosmology. The summary reiterates our major and minor discoveries.

*1. General state of astronomy in the 17th Century*

To supersede the Gregorian calendar with his own, Newton needed the tropical year to be well below $365^d\ 5^h\ 49^m$. Therefore, he selected the year ending in $48^m\ 45^s$, then changed it to another ending in $48^m\ 48^s$, but later computed it anew arriving at the third one ending in $48^m\ 57^s$. To explain why Newton twice changed his opinion regarding the tropical year we need to look into the state of 17th-Century astronomy.

*The tropical years found by 17th-Century astronomers*

Here are the tropical years found by the leading 17th-Century astronomers:
$365^d\ 5^h\ 48^m\ 45^s.5$ — Tycho Brahe (*Astronomiae instauratae progymnasmata*, 1602)
$365^d\ 5^h\ 48^m\ 57^s.6$ — Johannes Kepler (*Tabulae Rudolphinae*, 1627)
$365^d\ 5^h\ 49^m\ 4^s.5$ — Ismael Boulliau (*Astronomia Philolaica*, 1645)
$365^d\ 5^h\ 48^m\ 40^s$ — Fr. Giovanni Riccioli (*Almagestum Novum*, 1651)
$365^d\ 5^h\ 49^m\ 1^s$ — Thomas Streete (*Astronomia Carolina*, 1661)
$365^d\ 5^h\ 49^m\ 4^s.4$ — Vincent Wing (*Astronomia Britannica*, 1669)
$365^d\ 5^h\ 49^m$ — John Flamsteed (*De Inaequalitate Dierum Solarium*, 1672)

Newton did not own the first three books in the list. However, he may have learned about their tropical years from Riccioli's *Almagestum Novum*[7], which formed part of his library[8]. Newton also owned the books by all three English astronomers, to be discussed later.

The most striking feature of the list is the fairly large range, $25^s$, in the tropical year assumed by contemporary astronomers. The reason for such a disparity lies in the 17th-Century's confusion in two matters: suspicion of several disparate observations of the stars by various Greek astronomers reported by Ptolemy, and difficulty in integrating Ptolemy's equinox observations into their systems.

Indeed, each of the 17th-Century astronomers found the tropical year by comparing contemporary observations with those of ancient astronomers. They employed *two different* techniques: either a direct one — comparing the Sun's daytime crossings of the equinoctial points; or an indirect one — *via* estimating the precession of the equinoxes by comparing longitudes of the same star, both at two mutually distant epochs.

The former (direct) method was by far the more popular. Tycho Brahe compared his own equinox observation in 1588 with the one by Bernhard Walther in 1488. He did not use ancient equinoxes because the results were inconsistent[9]. Riccioli compared his own equinox observation of 1646 with one of Hipparchus of 159 BC[10]. Vincent Wing compared equinox observations of al-Battani (*c.* 882) and Ismael Boulliau (*c.* 1640)[11]. Kepler did not explain how he obtained the tropical year he uses in the *Tabulae Rudolphinae*.

Thomas Streete followed the latter (indirect) method, pioneered by Tycho, comparing Tycho's star observations (*c.* 1600) with those of Timocharis (*c.* 300 BC). He then dismissed the latter as unreliable in favour of the medieval *Persian tables* (composed *c.* 1115 AD, either by Omar Khayyam or by his disciple Abd al-Rahman al-Khazini).

Each method had its own difficulties not yet sorted out by 1700. Paradoxically, Streete's suspicion of the ancient observations of the stars stemmed from the disbelief in the variation of the obliquity of the ecliptic (inclination of the pole). In his 1661 *Astronomia Carolina* Streete wrote: "But the Equinoctial with the Poles thereof are fixed in the Earth and movable in the Heavens, as the Precession of the Equinox demonstrates: And the Inclination of the Equinoctial to the Ecliptick, or the distance of their poles is invariable and constant in all Ages, as by some select and more certain observations will easily appear."[12]. Thus Streete postulated that there is not and never had been a change in the obliquity of the ecliptic. Ptolemy quoted for Eratosthenes and Hipparchus the obliquity of the ecliptic as 23° 51′ 20″[13], while the contemporary observations pointed on average to 23° 30′[14]. The large, 21′ 20″, discrepancy led Streete to suspect the *quality* of ancient observations and thus of their instruments.

Streete proceeds with an example, saying that upon comparing Tycho's observations of Spica *c.* 1600 with those of Timocharis in 300 BC, one can see a 26° 24′ change in the star's longitude and hence a figure of 50″ for annual precession (which, as we know, is the correct value). However, Streete dismisses Timocharis' observations as suspect and instead compares Tycho's observations of Spica and the last star of Pegasus with the positions of those stars from the *Persian tables*, obtaining on average 48″ for the annual precession. Streete continues: "We have also considered many other Observations Old and New; but in regard the more ancient Astronomers were destitute of conventional Instruments (as is evident by the discrepancy of their observations and by their manifest error in the greatest declination of the sun) and because the error of 24′ in declination amounts at least to one whole degree in Longitude; we have therefore (for want of better observations) made choice of some such applications of the Moon and Planets to Fixt stars (related by Ptolemy) as have most probability of truth, and comparing them with some other, limited the constant Annual Precession of the Equinox 48″, the motion in 100 years 1° 20′, and the whole revolution thereof in 27 000 years"[15].

"The greatest declination of the sun" is the obliquity of the ecliptic. The "manifest error" in comparing ancient and modern is therefore 21′ 20″, which is not far from Streete's 24′. Streete's logic was as follows. Timocharis' alleged error of 24′ in declination led to the 1° error in the longitude of Spica. Divided by the 1900 years that had passed between Timocharis and Tycho, the latter gives a 2″ error per year for annual precession, which increases a tropical year by 48$^s$. But since Streete adopted quite a low value, 365$^d$ 6$^h$ 8$^m$ 30$^s$, for the sidereal year, he arrived at 365$^d$ 5$^h$ 49$^m$ 1$^s$ for the tropical year — only 15$^s$ greater than the true one.

However, Streete's logic was flawed: he presumptuously concluded that the ancient astronomers (Timocharis, Hipparchus) had poor-quality instruments and thus were bound to err in measuring declination and longitude of the stars. Streete had his reasons. Taking Hipparchus' value for the obliquity at face value one had to expect ~1′ decrease in the obliquity per century. Writing in 1661, Streete would expect to see 2′ difference between observations made in 1460 and 1660. However, the European observations for 200 years before his work and available to him, from Peuerbach to Hevelius, would not allow him to admit any changes in the obliquity.

Indeed, the sample of 22 European-based observations between years 1460 and 1661 gives the average obliquity of 23° 30′ at the average year 1568[14]. Now, the OLS, applied to the above sample, gives the intercept equal to the average and shows no decline in obliquity ($p$-value = 0·9), while the application of the simple linear regression that we discuss here greatly depends not only on the average but also on a particular observation one trusts most. Say, if Streete trusted Tycho most (23° 29′ 48″ in 1590) then the regression line would have a negative slope; while if Kepler (23° 30′ 30″ in 1627), a positive slope. The noise in observations came from 'fancies' the European astronomers entertained about refraction and the Sun's parallax for the winter solstice; *e.g.*, Tycho estimated the latter as 3′ while Riccioli as 30″[16]. Of course, taking into account observations by Arab astronomers of the 9th and 10th Centuries would certainly tilt the regression line down, but again Streete did not know how good their instruments were either.

Though Tycho, Kepler, and Riccioli did believe in the long-term variation of the obliquity, Newton did not have the books of the former two in his library and might not entertain a high opinion of Riccioli. However, Newton had both editions of Streete's *Astronomia Carolina* and had to consult the first (1661) edition working on his *Theory of the Moon's Motion* (as Streete's tables were highly esteemed by John Flamsteed) and is known to have read the second (1710) edition.

*Nutation as another source of uncertainty*

This mindset of everything being constant in the heavens could not survive after the appearance of *Principia*. In a letter to Newton in 1700 June–July, Astronomer Royal John Flamsteed speaks of "nutation of the axis": "That the Earth's Axis is not always inclined at the same Angle to the plane of the Ecliptik is a discovery wholly oweing to you and strongly proved in the 4$^{th}$ book of your Prin. Phil. Nat. Math. How much the alteration of this Angle or the Nutation of the Axis ought to be you have not yet shewed: and whether you have yet determined or no, I know not."[17].

However, this falls short of being able to single out and then estimate the phenomenon responsible for a 21′ 20″ change in the obliquity of ecliptic! Newton used the Latin verb *nutare*, meaning "to nod with your head", twice in all three editions of *Principia*, referring to a very small, almost unnoticeable 'nod' of the Earth's axis twice a year, due to the effect of the Sun. Newton never associated *nutare* with the Moon[18].

Neither can we ascribe to Flamsteed an advanced knowledge of the 18·6-year periodic variation in the obliquity of the ecliptic, now known as *nutation* of the Earth's axis. His letter to Newton shows that he observed this phenomenon only for half a cycle, from 1689 till 1700, not recognizing its cyclic nature. Identification of the nutation in the modern sense came later, generally attributed to James Bradley, in *c.* 1743. It causes an 18·6-year periodic variation of the obliquity with a maximum of 9″·2 from its mean value, and is unrelated to the long-term variation of the obliquity[19]. Analytical proof of nutation's existence, based on Newton's theory of gravitation, was demonstrated by Jean le Rond d'Alembert in 1749. Being unable to estimate nutation, Flamsteed felt that this phenomenon was spoiling the measurements of the tropical year.

Streete was partially right: the Greeks, including Hipparchus, erred in the inclination of the Earth's axis — but not by as much as 24′! Owing to the long-term variation in the obliquity of the ecliptic, the error, according to the modern estimates, was only *one third* as much. Quite surprisingly, a close estimate also follows as a by-product from Newton's analysis (see Section 5). The source for that error most likely was an incorrect position of Hipparchus' equatorial ring.

*Ptolemy's legacy*

Ptolemy's legacy warrants another historical digression. Robert Newton brought this topic to the centre of the modern study of ancient astronomy, charging the author of *Syntaxis* (later known as the *Almagest*) with outright forgery and arguing that the first to show the fallacy of Ptolemy's equinox observations was Jean-Baptiste-Joseph Delambre. The latter not only discarded them as untrue in his *Histoire de l'Astronomie Ancienne* (1817), but in his later work, *Histoire de l'Astronomie du Moyene Age* (1819), he proved that Ptolemy falsified the data to justify his own model[20].

However, that discovery has a much longer history. Already Tycho and his disciples viewed Ptolemy's observations as problematic. Longomontanus argued for forgery, while Kepler devised an interesting excuse for Ptolemy, that of an 'omitted day' by the Roman priests in the year 139 that supposedly confused the Alexandrian astronomer[21].

Kepler's 'excuse' recruited followers for more than a century; Euler seems to have been one of them. However, it was completely rebuffed by Tobias Mayer in a letter to Euler of 1753 August 22[22]. Interestingly, Mayer based his argument on the lunar eclipses cited by Ptolemy, saying that the length of the lunar month would be greatly compromised if Kepler's 'excuse' be accepted. Being first in explaining how Ptolemy adjusted the timings of the equinoxes to justify his lunar theory, Mayer was unaware that Ptolemy forged most of his lunar 'observations' as well[23].

Why did Newton ignore Ptolemy's equinoxes? Belenkiy & Vila Echagüe[6] conjectured that it was John Wallis who called Newton's attention to a problem with Ptolemy's equinox observations. However, Newton could have learned about this problem from Wallis indirectly — *via* Flamsteed, as the latter collaborated with Wallis in several 1670s' publications. Moreover, by 1670 this problem could have been a part of common knowledge, as the 1669 diatribe by Vincent Wing witnesses: "About the solar year and its magnitude I will speak here at large, which many ancient and recent astronomers found variable, using feeble principles, based in nothing else than in the doubtful observations of Ptolemy, who followed Hipparchus, that had established the length of the year exactly in $365^d\ 5^h\ 55^m\ 12^s$. But if we reject Ptolemy and compare the observations of Hipparchus, Albategnius and Walther of Nuremberg with those of Tycho Brahe (which are free from errors due to parallax and refraction) we find the tropical year length always the same, as that illustrius Tycho and Longomontanus in *Astronomia Danica*, Book I, Chapter 5 Theorica cleverly uphold, assenting to them Johannes Kepler in *Epitome Astronomia Copernicana* page 927, where it is confirmed that the length of the year is the same from the time of Hipparchus, with the only exception of Ptolemy, against whom the observations of Hipparchus, Proclus, Albategnius and even the very learned Bullialdus agree, and our own restitution of the mean motion of the Sun confirms."[24].

*Why couldn't Newton just rely on contemporary observations?*

Desiring to reduce the uncertainty in the tropical year, Newton faced the fact that contemporary science lacked a reliable theory for the precession of the equinoxes. The first (1687) edition of *Principia* explains precession only *qualitatively*, not quantitatively, though it fully recognizes the Moon as the major force for the precession[25]. In 1694–5, Newton collaborated with Astronomer Royal John Flamsteed in an attempt to advance a theory of the Moon's motion. His employment with the Royal Mint in 1696 temporarily suspended those efforts. However, two years later Newton decided to make another attempt. A memorandum of David Gregory, supposedly of 1698 July 7, says: "On account of Flamsteed's irascibility the theory of the Moon will not be brought to a conclusion, nor will be any mention of Flamsteed, nevertheless he [Newton] will complete to within *four* [arc]minutes what he would have completed to *two*, had Flamsteed supplied his observations"[26]. Two months after the request of the Royal Society, in 1700 April, Newton penned a response to Leibniz, correcting Kepler's tables for the mean positions of the Sun and Moon. However, the precession of the equinoxes, a key component to compute the tropical year, seems to have baffled him long before that. Gregory's 1698 memorandum continues: "He [Newton] constructs afresh the whole theory of comets, and the precession of the equinoxes."[26].

Let us see why in 1700 Newton could not just copy the tropical year from Flamsteed. Precision of measurements of celestial objects depends greatly on the technical characteristics of the astronomical instruments. Though seriously improved since Tycho Brahe's time, the quality of the instruments still led to substantial errors, *too large* for Newton's purpose. For example, the instruments of two leading English astronomers at the end of the 17th Century, both Newton's close associates, Edmund Halley and John Flamsteed, were able to measure the Sun's position up to $5''$[27], and Newton certainly knew that fact. Though 17th-Century astronomy was ignorant of aberration and *true* nutation, atmospheric refraction was under discussion. Newton was well aware of it: he discussed refraction of light near the horizon in an extensive letter exchange with Flamsteed in 1694[28]. Inability to estimate the atmospheric refraction, together with Flamsteed's "nutation of the axis", could increase a possible error in a single observation to $10''$.

Being ignorant of the calculus of errors, Newton could intuitively stick to the latter value. (Modern statistics would support his intuition as well: two observations were needed to find the Sun's transit over the celestial equator; whatever was the possible error for one, their linear interpolation would reduce a possible error by $\sqrt{2}$; comparison of two equinoxes would increase a possible error by $\sqrt{2}$, back to the original value.) Since the Sun's daily motion near the equinoxes in declination is 24′, the 10″ accuracy was tantamount to an error of 10 minutes, and this was indeed the case for Flamsteed.

To prove this, we computed the time of the 1681 vernal and autumnal equinoxes from Kollerstrom's spreadsheet for Flamsteed[29] and compared it with the modern formula[30]. The results are expressed in Gregorian dates and Universal Time ($\Delta T$ for 1681 was only 1 sec[31]):

Vernal equinox (Flamsteed): 1681 Mar 19, 23$^h$ 35$^m$; modern estimate: 1681 Mar 20, 0$^h$ 0$^m$.

Autumnal equinox (Flamsteed): 1681 Sept 22, 12$^h$ 09$^m$; modern estimate: 1681 Sept 22, 12$^h$ 02$^m$.

The differences are −25 min and +7 min, respectively, consistent with the above estimate.

Even employing Flamsteed's equinox observations at Greenwich 20 years apart, say, of 1681 and 1700, would not reduce the uncertainty in the tropical year below half a minute. Sensing that Flamsteed's observations alone could not suffice, Newton looked for different tools.

*2. Newton's quest for the tropical year*

Newton redrafted his calendar proposal three times, each time citing a different tropical year.

*The year ending in 48$^m$ 45$^s$*

Belenkiy & Vila Echagüe[6] proposed that when writing the first draft, with a tropical year of 365$^d$ 5$^h$ 48$^m$ 45$^s$, Newton first picked an arbitrary book from the shelf, which happened to be that of Tycho Brahe. That conjecture seemed problematic, since Newton did not own any book by Tycho, except possibly the one on comets. However, it was further supported by the fact that *Institutionum Astronomicarum Libri Duo* (1676) by Nicolas Mercator, well read by Newton[32], does contain Tycho's tables in the appendix of that book, and it was an easy matter for Newton to derive the tropical year from them. Still it remained unclear why Newton picked Mercator's book from the shelf and chose Tycho's year, which, moreover, was not written explicitly.

*The year ending in 48$^m$ 48$^s$*

The origin of the year ending in 48$^m$ 48$^s$ remained a puzzle even longer. At the time, when the 17th-Century texts were practically unavailable, Belenkiy & Vila Echagüe[33] hypothesized that it could have come from Vincent Wing or Thomas Streete. Indeed, the author of *Principia* keenly followed the works by both astronomers over many years. For example, Newton spotted and corrected the error for the precession of the equinoxes in his own copy of the 2nd edition of *Astronomia Carolina* (1710), which shows his close familiarity with the first edition (1661)[34]. Besides, working on determination of the day of the Crucifixion, Newton used data from Wing's 1669 *Astronomia Britannica*[35].

However, our hypothesis was soon rejected after we discovered that Newton had got hold of *Astronomia Carolina* and *Astronomia Britannica*.

The true solution was pointed out to us by an anonymous referee. The year ending in 48$^m$ 48$^s$ belongs to Giovanni Battista Riccioli and can be found on page 16 of Riccioli's *Astronomia Reformata* (1665). This was unexpected since Newton did not own that book at his death. The only book of Riccioli he owned was *Almagestum Novum*, which had the tropical year ending in 48$^m$ 40$^s$. Thus,

Newton either borrowed *Astronomia Reformata* from someone else or owned it in 1700 but later sold it. But why would he look into Riccioli's book in the first place?

The answer emerges from the examination of the table from *Yahuda MS 24 D*, displayed in Fig. 1, with the dates of Hipparchus' autumnal equinoxes in years 162–159–158 BC and 147–146–143 BC and vernal equinoxes in years 146–135–128 BC[36]. As it turns out, it was copied from Flamsteed's table from *De Inaequalitate Dierum Solarium* (1672)[37] reprinted in a 1678 collection of essays by Jeremiah Horrocks and John Wallis. In his turn, Flamsteed also copied the dates in the left part of his table (see Fig. 2) from Riccioli's *Astronomia Reformata*[38].

True, Flamsteed not only copied Riccioli's data but also made a step forward. After translating the timings of Hipparchus' equinoxes into 'Derby Time' (1°·5 to the west of Greenwich), he computed the mean positions of the Sun at those moments assuming the tropical year of $365^d\ 5^h\ 49^m$ and the equation of time of $8^m$. It seems Newton could use those ready data to find a correction to Flamsteed's year. Indeed, the autumnal set is off the mark on average by 17′, the vernal by −10′, and the averaging would result in practically the same correction, −3 sec, that Newton obtained the hard way described below, in Section 4. Why then did he choose the hard way?

The poor estimate of Alexandria's longitude assumed by Flamsteed (most likely computed from several entries in *Almagestum Novum*) was inessential and easily amendable, but Newton could have felt that the result might be different with a different set of parameters for the Sun's motion. From the mid-1690s exchange with Flamsteed he learned that the Astronomer Royal evaluated several solar parameters more precisely than in 1672 in *De Inaequalitate*. That's why parameters of *Yahuda MS 24 D*, those for the mean positions and mean motion of the Sun as well as the position of the solar apogee, Newton borrowed from Flamsteed's later work, *The Doctrine of the Sphere* (1680), published in a 1681 collection of essays under the editorship of Jonas Moore[39].

Fig. 1

*Yahuda MS 24 D*, the upper third. Hipparchus' sample of ten equinoxes quoted in Ptolemy's *Almagest*. Column 1 specifies the year's number within the Third Callipic Cycle of 76 years. Column 2 translates them into proleptic Julian BC years. The last column shows Newton's first attempt to convert Alexandrian Time to Greenwich Time assuming a $2^h\ 16^m$ difference.

Fig. 2

The table of Hipparchus' equinoxes from Flamsteed's *De Inaequalitate* (1672). There are some irregularities in the second column (136 must be 146; 128 is missing in the last line). The timings were translated into 'Derby Time' — $2^h 50^m$ from Alexandria for the autumnal set and $2^h 34^m$ for the vernal. (The difference is due to the equation of time.) In the last column are the mean positions of the Sun.

Though *The Doctrine* became the primary source for *Yahuda MS 24 D*, the parameters he used might rather have come from an extensive letter exchange with Flamsteed of the mid-1690s. Indeed, the maximal equation of the centre, given as 1° 59′ in *De Inaequalitate* and 1° 55′ in *The Doctrine*, was changed in *Yahuda MS 24 D* to 1° 56′ 20″ (as we found by solving Kepler's equation), known to be found by Flamsteed in 1692.

Newton certainly spotted several typographical errors in timings for Hipparchus' vernal-equinox set and wanted to clarify the status of two observations of 146 BC. Since he did not own a copy of *Almagest*, at that point he had to consult Flamsteed's source, *Astronomia Reformata*. That is how Newton arrived at the second year, that of Riccioli, and, most likely, the first one as well, since Riccioli cites there the years of other astronomers too, in particular, Tycho. The new calendar Newton envisaged had the average year ending in $48^m 46^s$, so his first two choices were quite obvious!

## The year ending in $48^m 57^s$

Newton arrived at the year of $365^d 5^h 48^m 57^s$ after two pages of intricate computations. How accurate that value is depends on the units of time with which one expresses the tropical year. Newton was not aware of the fact that the tropical year and the mean day change with time. If we take the unit of time to be the ephemeris second, related to the tropical year 1900, the tropical year varies according to formula $(31556925 \cdot 9747 - 0 \cdot 530\ T)se$, where $se$ is the ephemeris second defined for 1900 and $T$ is number of centuries since 1900[40]. This formula gives $365^d 5^h 48^m 57^{se}$ for Hipparchus' time (160–130 BC), and $10^{se}$ less for Newton's own time, 1700. The average value between the two lies at $365^d 5^h 48^m 52^{se}$. With that, Newton seems to make just a $5^{se}$ mistake.

Neither Newton nor Hipparchus measured the tropical year in ephemeris seconds, however; rather, they used *seconds of their time*, "mean seconds", computed as $1/(24 \times 3600)$ of the mean day. Dividing the above-mentioned formula for the tropical year by the length of the mean day $(86400 + 0·0015\,T)$ *se*, we computed the tropical year for years 1700 AD and 160 BC expressed in "mean seconds" (*mse*). The results are $365^d\ 5^h\ 48^m\ 48^{mse}$, for 1700 AD, and $365^d\ 5^h\ 49^m\ 8^{mse}$, for 160 BC. The average of the two is $365^d\ 5^h\ 48^m\ 58^{mse}$ — again quite close to Newton's value.

*3. The Ordinary Least-Squares Method*

Since we claim Newton's method was a *regression analysis*, a clarification is due. Indeed, the regression analysis is usually identified with the Ordinary Least-Squares (OLS) regression and algebraic *minimization* ideology. But in the problem where a slope of the regression line is sought, the geometric intuition suggests a simpler model. To show where this model stands in relation to the OLS method, let us make an historical digression.

Linear-regression analysis aims at the approximation of data, represented by a set of points $(X_n, Y_n)$ on the X,Y plane, by a single straight line. The OLS method claims that there is a unique line $Y = \hat{a} + \hat{\beta}X$ such that the sum of the squared distances from every point to this line is *minimal*. Further, assuming that the data $(X_n, Y_n)$ are scattered around a certain line randomly, with zero mean and equal variance, the above $\hat{\beta}$ is the *best linear unbiased* estimator. The 'best' here means that $\hat{\beta}$ is the *most effective*, or has the *least variance* (the notion that Newton did not know) among all unbiased estimators.

Notice that if the slope $\beta$ is initially assumed to be zero, the minimization leads to a horizontal line $Y = \hat{a}$, where ('the best') $\hat{a}$ is $\bar{Y}$ — the average over all *Y*s. The latter gives the least variance $\Sigma_n (Y_n - a_n)^2$ among all possible $a$'s. Remarkably, even the most eminent of the early 17th-Century astronomers, Galileo and Kepler, never actually grasped the latter property of the averages and never used them[41]. Several examples should illustrate this.

Anders Hald cites Galileo regarding the errors in observations of the new star of 1572 as saying in *Dialogo* that "the most probable hypothesis is the one which requires the smallest corrections of the observations", and further suggests that Galileo used *the sum of the absolute deviations from the hypothetical value* as his criterion[42]. But, as we presently know, this criterion points rather to the *median* of a sample, not to its *mean* (average). This is consistent with Galileo's willingness to consider even "impossible results" or the so-called *outliers*, *i.e.*, the observations that are numerically distant from the rest of the data[43], since the median is actually independent of the outliers while the mean is strongly dependent on each of them.

There was a long-time conviction that at least on one occasion Tycho used the *mean* of a sample[44]. However, the quick analysis of the data implies that the value 26° 0′ 30″, which Tycho finally chose from a sample of 15 observations of the right ascension of the star α Arietis, was *not* the mean but the *median* of the sample.

In the general case, with non-zero $\beta$, the most popular regression method is OLS, which finds first the differences (residuals) $\hat{u}_n$ between the data and the regression line and then searches for the pair $(\hat{a}, \hat{\beta})$ that minimizes the total sum of squares of the residuals, $\Sigma_n \hat{u}^2_n = \Sigma_n (Y_n - \hat{a} - \hat{\beta} \cdot X_n)^2$. Discovery of the OLS regression belongs to Adrien-Marie Legendre (1805) while Carl Friedrich Gauss later (1809, 1821) provided a probabilistic setting for the theory[45].

Minimization over each parameter of the above pair leads to the two following equations:

$$\Sigma_n u_n = 0 \tag{1}$$

and

$$\Sigma_n u_n X_n = 0 \tag{2}$$

Equation (1) is equivalent to the fact that the regression line passes through the point $\bar{X}, \bar{Y}$, and we shall see that Newton accomplished exactly that, finding an average time of the equinox in the 'average' year of the set. Of course, one equation is not enough to find two parameters, therefore Newton put the less interesting of the two, the intercept $\hat{a}$, equal to zero. The second equation (2), which leads to the OLS regression, was missed by Newton.

Let us translate the problem Newton faced concerning the X,Y reference system. Actually, he had to combine two different frameworks into one. If $Y_n$ is daytime of the equinox in a Julian year $X_n$, as given by Hipparchus, then the coefficient $\hat{\beta}$ represents a 'deficiency' of the tropical year *vs.* the Julian year.

If $Y_n$ is the position of the Sun on the ecliptic at the calendar time $X_n$ (expressed in Julian years), as given by Flamsteed, then the coefficient $\hat{\beta}$ represents a shift of the equinoxes along the ecliptic against the Sun's motion for one Julian year. Together with a (known) sidereal year, this shift of the equinoxes provides a tropical year. In both cases the resulting estimator $\hat{\beta}$ must be negative: ~ –11¼ minutes per year for Hipparchus, and ~ –27″ per year for Flamsteed. The latter result, however, can be easily translated into the former. A four-year Julian cycle poses a difficulty due to a leap day. To circumvent the difficulty, Newton moved all the equinoxes to the neighbouring Julian years divisible by 4 (proleptic leap years), re-scaling the coefficient $\hat{\beta}$ by factor of 4.

The weakness of a straightforward application of *any* regression technique to Hipparchus' data is obvious: the sample of *ten* observations is too *thin*. Applying the OLS method to Hipparchus' data, the authors found the regression coefficient $\hat{\beta}$ far from the expected value of –11¼ minutes — the second trio of the autumnal equinoxes appeared to be the major culprit. However, Hipparchus' sample can still be used if several faraway observations are added to it. This was Newton's first insight: to add Flamsteed's two observations to Hipparchus' ten.

## 4. Newton's linear-regression model

Newton had two remarkable insights, first separating qualitatively different observations into two groups, and secondly choosing an estimator with several good properties. We shall follow his argument closely, explaining the difficulties he faced and the ways he circumvented them. His way into the unknown can be traced by analyzing the corrections he made at several junctions on his journey.

*Description of Newton's procedure*

Newton could not have computed an 'honest' average time of the equinoxes at the 'honest' average year, since the latter would be a fraction. Therefore, he applied the following procedure (see Fig. 3).

First Newton separated autumnal equinoxes from vernal (column 1 on the left designates the proleptic Julian BC years) and arranged Hipparchus' equinoxes chronologically for the *two* groups separately (column 3). Next he chose two 'anchor' equinoxes: 158 BC Sept. 27, 0:00, for the autumnal group and 135 BC March 23, 12:00, for the vernal group (both in Alexandrian local time). Taking

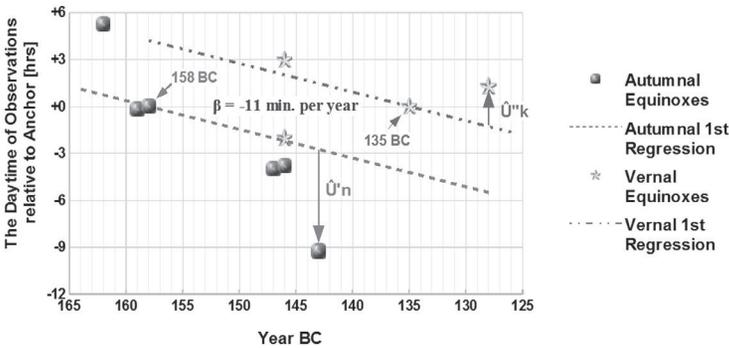

Fig. 3

The middle part of *Yahuda MS 24 D*: proleptic Julian years in column 1; the timings of Hipparchus' nine equinoxes in column 3; regression lines with Flamsteed's initial slope in columns 2, 6, 7.

'anchors' as given, he linearly extrapolated, separately for each of two groups, the time of day when the rest of the equinoxes had to occur, assuming that the year of $365^d\ 5^h\ 49^m$ is correct (column 2). These are two regression lines, though chosen somewhat arbitrarily, with the slope $\beta = -44^m$ per 4 years.

Next, Newton subtracted column 2 from column 3, thus finding (column 4) how far Hipparchus' sample lies off the regression lines. He got: $-5^h\ 16^m$ for 162 BC, $+11^m$ for 159 BC, 0 for 158 BC, $+3^h\ 59^m$ for 147 BC, $3^h\ 48^m$ for 146 BC, and $+9^h\ 15^m$ for 143 BC in the autumnal group; $-29^m$ for 146 BC, 0 for 135 BC, and $-1^h\ 17^m$ for 128 BC in the vernal group. These are the residuals $u_n$ shown in Fig. 4, where the icons represent ten of Hipparchus' equinoxes. The balls are the six autumnal equinoxes of 162, 159, 158, 147, 146, and 143 BC, while the stars are the four vernal equinoxes of 146 (two), 135, and 128 BC.

Fig. 4

Residuals of Hipparchus' equinoxes computed by Newton from the regression lines he drew *via* the 'anchor' equinoxes with the slope $\beta = -11$ minutes (the difference between Flamsteed's and the Julian year).

Then, Newton summed up the residuals to $12^h\,29^m$ for autumnal and $-1^h\,46^m$ for vernal equinoxes, and finally averaged the residuals to $2^h\,5^m$ and $-35\tfrac{2}{3}^m$ for the two groups, respectively (two marginal marks in column 3), finding how far off his arbitrarily chosen regression line passes from the 'average point' $(\bar{X},\bar{Y})$.

As a side remark let us note that Newton made a tiny error. The correct value for the autumnal set is $1^h\,59^m\,30^s$. When adding up all the values to compute the average, instead of subtracting $5^h\,16^m$ for year 162 BC, he subtracted 5 hours but added 16 minutes. The 32-minute difference, divided by 6, leads to the 5·4-minute difference in the average. This changes his result by 13″.

Newton did not choose the right strategy at once. First he decided to move all samples up, each group by its average (column 5); however, realizing that this was the wrong move, he crossed the column and moved both regression lines down by the respective average (column 6). This is equivalent to making the sum of the residuals equal to zero and hence the imposition on the regression line of the first of the two 'normal equations' (eq. 1) — see Fig. 5.

The first normal equation is tantamount to the condition that the regression line must pass through the point $(\bar{X},\bar{Y})$:

$$Y - \bar{Y} = \hat{\beta}(X - \bar{X}). \qquad (3)$$

Next, Newton converted Alexandrian local time for the points on the regression line into Greenwich local time by subtracting from column 6 the $2^h\,15^m$ difference in longitude between the two locations (column 7). In particular, the 'anchor' equinoxes were set out 158 BC Sept 26, 21:55, and 135 BC March 23, 12:35, (19:40 and 10:20 Greenwich local time). At this point, Newton, for unclear reasons, reassigned the vernal 'anchor' equinox to the one of 146 BC March 23, 20:36 (18:21 Greenwich local time).

At that point Newton made a slip of the pen. Computing the Greenwich Time for the autumnal 143 BC equinox, he subtracted $3^h\,15^m$ instead of $2^h\,15^m$. However, that slip did not influence any conclusions since he did not use that equinox again. Finally, Newton placed two new points on the graph: A (158 BC Sept. 26, 19:40) and V (146 BC March 23, 18:21), and drew through them two new regression lines, parallel to the old ones (see Fig. 5).

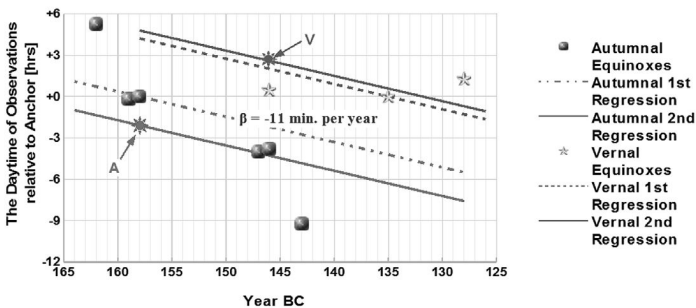

Fig. 5

New regression lines adjusted for the first 'normal' equation (eq. 1). The anchor vernal equinox was changed from the one in 135 to the one in 146 BC.

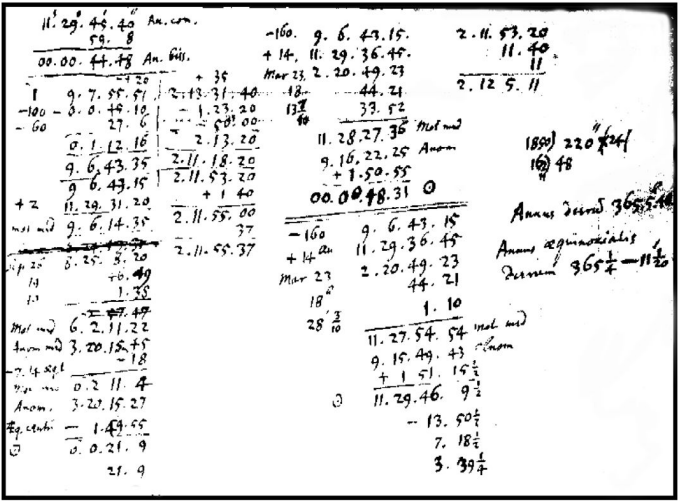

Fig. 6

*Yahuda MS 24 D*: Newton's correction of Flamsteed's tropical year: in the first two columns the Sun's position at the 'anchor' autumnal equinox of 158 BC, in the last two — at the 'anchor' vernal equinox of 146 BC, followed by their averaging at the bottom. The concluding computation of the year is in the margin.

Newton further computed the Sun's mean longitude for both dates, 158 BC Sept. 26, 19:40, and BC 146 March 23, 10:20, by using the mean Sun's positions identical to those of Flamsteed's data from *The Doctrine*. In the first three lines above column 1 (see Fig. 6), he computed the Sun's mean motion for a year of 365 days as $11^s$ 29° 49′ 40″ and for a year of 366 days as 44′ 48″ *mod* 360°. Note that Newton used the superscript letter $^s$ to designate a 'sign' (30°, equivalent to an average zodiacal constellation), thus $11^s$ 29° 49′ 40″ = 359° 49′ 40″.

The starting point for his computations is the mean Sun's position at $9^s$ 7° 55′ 51″ for 1 AD Dec. 31, noon, computed backward from Flamsteed's 1680 data. However, in the process of computation, Newton corrected the mean Sun's position by –20″ and the solar apogee's position by 35′. These corrections were communicated to him by Flamsteed in the mid-1690s[46]. In the short columns 2 and 4, Newton computed the positions of the Sun's apogee. They were needed for the computation of the 'equation of the centre' (Kepler's equation) in columns 1 and 3. In column 2, the Sun's apogee on 158 BC Sept. 26 was found at $2^s$ 11° 55′ 37″. Above column 2 is a later correction of 35′ for the solar apogee's tabulated position. In column 4, the Sun's apogee on 146 BC March 23, was found at $2^s$ 12° 5′ 11″. Surprisingly, while borrowing from Flamsteed the contemporary solar apogee's position, Newton mistakenly equated the apogee's motion with the precession of the equinoxes, 1° 23′ 20″ per century! We shall discuss this point later.

In columns 1 and 3, Newton computed the mean and then the true positions of the Sun for his anchor equinoxes. In column 1, he computed the Sun's mean longitude, at 19:40 on 158 BC Sept. 26, at $6^s$ 2° 11′ 22″, and found the anomaly of $3^s$ 20° 15′ 45″ (by subtracting the position of the solar apogee in column 2 from the mean longitude). Only at this point did he decide to switch from the *apparent* time to the *mean* time. The difference, the so-called 'equation of time', was $7^m\ 14^s$, and he had to reduce the mean longitude and anomaly by $18''$, to $6^s\ 2°\ 11'\ 04''$ and $3^s\ 20°\ 15'\ 27''$, respectively, before computing from the latter the 'equation of the centre' as $-1°\ 49'\ 55''$. (He did not give details of the last computation — it is likely that he gleaned the result from Flamsteed's tables.) Finally, he found the Sun's true longitude at the anchor autumnal equinox as $6^s\ 0°\ 21'\ 9''$, *i.e.*, $+21'\ 9''$ off $180°$.

For the anchor vernal equinox of 146 BC March 23, 10:20, Newton computed the reduction from apparent to mean time first, *subtracting* the equation of time ($7^m\ 14^s$) from $18^h\ 21^m$ to obtain $18^h\ 13^m\cdot 7$. He then computed the Sun's true longitude, getting a (wrong) position at $+18'\ 31''$ off $0°$. At this point, he noticed his mistake. Near the two equinoxes, vernal and autumnal, the equation of time has the same absolute value, but *opposite signs*! Newton separated the wrong computations by a double line and repeated his calculations, this time *adding* $7^m\ 14^s$ to $18^h\ 21^m$ and arriving at $18^h\ 28^m\cdot 3$. The Sun's mean longitude at 10:20 on 146 BC March 23 was $11^s\ 27°\ 54'\ 54''$, the mean anomaly was $9^s\ 15°\ 49'\ 43''$, and the equation of the centre was $+1°\ 51'\ 15\frac{1}{2}''$. Finally, the Sun's true position was found at $11^s\ 29°\ 46'\ 9\frac{1}{2}''$ or $-13'\ 50\frac{1}{2}''$ off $0°$.

Therefore, Newton found the *asymmetry* of $34'\ 59\frac{1}{2}''$ in the motion of two equinoxes. He did not even consider a solution in which both equinoxes, autumnal and vernal, move with different speeds. Instead, Newton tacitly made a step which he did not explain in writing. To compensate for this *bias*, accumulated over 1850 years, he divided the asymmetry in half, to $17'\ 29\frac{3}{4}''$, and then moved Hipparchus' vernal anchor equinox *forward* and his autumnal anchor equinox *backward* by that amount to the new positions, such that both equinoxes then appeared to be ahead of their true, $0°$ and $180°$, positions by the same arc of $3'\ 39\frac{1}{4}''$.

The way to resolve the remaining problem was clear: to increase the speed of the equinoxes by $3'\ 39\frac{1}{4}''$ (divided by 1850 years), or, equivalently, decrease the year by 3 seconds *via* equation:

$$3'\ 39\frac{1}{4}'' \times 24\ (s/'')\ /\ 1850\ y = 3\ s/y, \qquad (4)$$

where $24\ (s/'')$ is the inverse speed of the mean Sun, *i.e.*, the time during which the mean Sun moves one arc-second.

In this way, Newton found the *annus equinoxialis* as $356^d\ 5^h\ 48^m\ 57^s$.

*Properties of Newton's estimator*

Taking any position of the Sun tabulated by Flamsteed as $(X, Y) = (0,0)$, the point is equivalent to fixing $a = 0$ in the linear regression model (see Fig. 7). With that, Newton's estimator essentially is the slope of a regression line through the centre $(0,0)$ and the point $(\bar{X}, \bar{Y})$:

$$\beta\dagger = \bar{Y}/\bar{X}, \qquad (5)$$

which is obtained by putting $X = 0$, $Y = 0$ in equation (3).

This is a well-known estimator in the field of regression analysis, though rarely cited[47]. Actually, it is the simplest estimator (usually discovered by college students in the first year!) and therefore, historically, must be discovered *first*, earlier than a sophisticated OLS estimator. From this point of view, our paper establishes 'historical justice', finding this method being discovered — though not published — before discovery of the OLS.

The problem with the estimator $\beta\dagger$ is that it might be biased if $a \neq 0$. But certainly Newton had serious reasons to consider $a = 0$ as a true situation! As

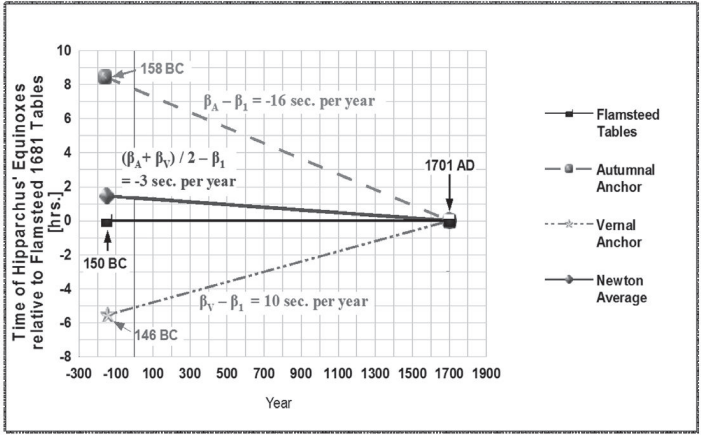

Fig. 7

Newton's final adjustment of the regression slope by averaging of the two slopes that were separately derived for autumnal and vernal sets of the equinoxes.

we pointed out, Flamsteed's error in the position of the equinox in 1681 could hardly be greater than 10$^m$, while Hipparchus erred by as much as 5 or even 9 hours! Therefore Newton could have viewed Flamsteed's equinox observations he used as 'precise' — which was tantamount to $a = 0$.

The grossly diverse sizes of the errors for different observations invalidate assumptions of the second part of the Gauss theorem and, therefore, one cannot readily dismiss estimator $β†$ as *less efficient* than the OLS estimator $\hat{β}$. Yet Newton was not aware of the notion of variance and did not know the modern (so-called 'weighted') method to handle the problem of *heteroscedasticity*.

*Newton's algorithm*

Formally, Newton's algorithm can be reformulated as follows:

0. Choose two 'anchor' equinoxes: autumnal in year $X_A$ and vernal in year $X_V$.
   Define: $σ = 24$ (s/″) = inverse of the Sun's speed and
   $T = (158 + 146)/2 + 1700 \cong 1850y$.

1. Choose $β_1 = 365^d\ 5^h\ 49^m - 365^d\ 6^h = -11$(min/y). (Fig. 3)

2. Choose two 'anchor' equinoxes A and V. Form a regression line with given $β$ from the anchors (Fig. 4).

3. Using the first normal equation (eq. 1), compute an average time of the anchor equinoxes $\bar{Y}_A$ and $\bar{Y}_V$ in the years $X_A$ and $X_V$. (Fig. 5)

4. Use Flamsteed's mean equinoxes $Y_A^F$ and $Y_V^F$ for year $X_0 = 1701$.
   Compute $β_A$ and $β_V$ by formula (3): $β_A = \dfrac{\bar{Y}_A - Y_A^F}{X_A - X_A^F}$ and, similarly $\dfrac{\bar{Y}_V - Y_V^F}{X_V - X_V^F}$.

5. Find $β_2 = \dfrac{β_A + β_V}{2}$.

6. Return to step 2.

Note that in step 4 Newton actually found
$\frac{T}{\sigma}\beta_A = \frac{T}{\sigma}\beta_1 - 21' 09''$ and $\frac{T}{\sigma}\beta_V = \frac{T}{\sigma}\beta_1 + 13' 50''\cdot 5$.

Averaging these two quantities up in step 5 he obtained
$\beta_2 = \beta_1 - 3' 39''\cdot 25 \frac{\sigma}{T} = \beta_1 - 3^s$. (Fig. 7)

*Iterations*

Iterations (Step 5) are necessary since the anchor equinoxes are not the points $(\bar{X}, \bar{Y})$. The final line(s) would approximate the one(s) that passes through the latter point(s). That Newton recognized this fact and was ready for iterations follows from his hesitation in draft A3 over the final value: $56^s$ or $57^s$.

This procedure seems intermittently iterative because of nonlinearities from the application of Kepler's equation and the equation of time in computing the 'average' time of both equinoxes in Step 2. After correcting $\beta$, the non-linear terms must produce the second-order corrections and the whole procedure must be repeated again. However, this procedure would converge very quickly, since nonlinearities mutually compensate for each other. For example, the *equation of the centre* for the Sun near the autumnal equinox was $-1° 49' 55''$, while near the vernal it was $+1° 51' 15''$. The same situation occurred for the *equation of time*: $+7^m\cdot 3$ near the vernal equinox and $-7^m 14^s$ near the autumnal one. As a result, Newton did not bother to perform the second step of iteration.

## 5. Newton's astronomical worldview

*Proper motion of the apogee*

It seems that Newton did not believe in the *proper motion* of the solar apogee since he used for it the same value as that for precession of the equinoxes (1° 23′ 20″per century). The reason may be traced to his remark in the *Principia*, where he wrote, in reference to the planets: "The aphelia and the nodes don't move."[48] He meant that the other planets and the comets produce negligible effects on the Earth. In contrast, Flamsteed computed the apogee's motion as 1° 45′ per century or 1′ 3″ per year, the same as Kepler found almost a century earlier[49].

Newton was certainly influenced by Thomas Streete who wrote (with a later reference to Tycho): "The Aphelions and Foci of the Middle motions of the Primary Planets are (as well as the Centers of the Sun and Fixt Starres) Immovable, the augmentation of their Longitudes being only the Præcession of the Æquinox; and that is not barely the opinion of my selfe & some others; the Observations which shall be produced in their convenient place will sufficiently demonstrate; and hence the Sidereall years are alwaies equall, but the Tropical yeares unequall."[50].

Still, the true position of the apogee troubled Newton, and in the middle of his computations on the regression model he added a 35′ correction to the apogee's position suggested by Flamsteed. This addition did not change much: the final position of the apogee in 146–158 BC found by Newton (~ 72°) is quite far from what Hipparchus claimed (65° 30′).

If Newton had believed Flamsteed in that matter, he would have placed the solar apogee at 65° 17′ 32″ in 158 BC and at 65° 29′ 36″ in 146 BC. With that, the solar anomaly, the equation of the centre, and the true position of the Sun in 158 BC would be, respectively, 116° 53′ 32″, −1° 44′ 44″, and 180° 26′ 21″, while in 146 BC, respectively, 292° 25′ 17″, 1° 46′ 39″, and 359° 41′ 33″. The asymmetry would be 44′ 48″, while the correction to the year would be 3′ 57″ or 3$^s$·07, slightly greater than that found earlier. This could be another good reason for Newton's hesitation in choosing between an ending of 57$^s$ or 56$^s$ for the tropical year in the third draft.

Another remark is of some historical interest. Though contemporary astronomers would fix the starry sky (and/or apogee) in space, allowing the equinoxes to move, Newton's computation reveals that he held — at least for computational purposes — the older view and kept the equinoxes fixed, allowing the sky and apogee to move.

*Data arrangement*

The vernal equinox of 146 BC March 23, for which Hipparchus quoted two diverse reports, one from Rhodes and one from Alexandria, at 18$^h$ and 23$^h$, respectively, gave much trouble to later historians of science[51]. On the contrary, Newton considered both data as valid and assumed the *average* timing, 20:30, as the time of the equinox. From the orthodox point of view, those two observations received 'half weight' compared with two other vernal equinoxes in 135 BC and 128 BC. But it is completely justifiable in this situation, otherwise one year, 146 BC, would become twice as important. Stigler writes: "Astronomers averaged measurements they considered to be equivalent, observations they felt were of equal intrinsic accuracy because measurements had been made by the same observer, at the same time, in the same place, with the same instrument and so forth. Exceptions, instances in which measurements not considered to be of equivalent accuracy were combined, were rare before 1750."[52].

As we have seen, such a rare event occurred as early as 1700. Our guess is that Roger Cotes, editor of the 2nd edition of *Principia*, who formulated a similar thought[53] *c.* 1716, learned it from his letter exchange with Newton in 1712–13. That could be a subject of further inquiry.

*Reflection on Hipparchus*

Moving Hipparchus' anchor vernal equinox forward, while the autumnal one backward, Newton tacitly assumed that Hipparchus' 0° and 180°+ marks were displaced by 17′ 29¾″ in longitude — though in *opposite* directions. Multiplying this value by sin 23°·5 (= 0·4) converts it into the 7′ error in declination. Thus, Newton's premise could have been that Hipparchus' equatorial ring was *biassed* down by 7′. This observation follows directly from Flamsteed's table in Fig. 2 as well.

To explain the bias of Hipparchus' ring, Robert Newton suggested that Hipparchus never conducted an observation of a winter solstice but only of a summer solstice. Then he indiscriminately assumed the obliquity found by Eratosthenes, 23° 51′ 20″, which exceeded the true value by 8′. Thus, his equatorial ring must be biassed down by the same amount[54].

True or not, modern estimates by Simon Newcomb and Robert Newton claim ~ 23° 43′ 20″ for the *inclination of the pole* (obliquity of the ecliptic) at Hipparchus' time, thus assigning to Hipparchus an 8′ (±3′) error in the obliquity[55].

Interestingly, Euler touched on this very subject in his letter to Mayer of 1755 May 27: "The obliquity of the ecliptic decreases by about 48″ per century… According to this it is quite certain that the obliquity of the ecliptic has previously been larger than now, yet it could have been nowhere near 23° 51′ at the time of Hipparchus."[56]. Since in 1756 Mayer found the obliquity to be 23° 28′ 16″, he would add the 15′ 12″ change during 19 centuries between him and Hipparchus, to arrive at 23° 43′ 28″ — in excellent agreement with both Newton's and modern estimates. In the letter, Euler adds that Mayer may "undoubtedly discover the reason for the error by himself." Unfortunately this is the very last letter of their exchange translated by Forbes, so we don't know Mayer's answer, if any.

The issue was raised again in 1982 by Dennis Rawlins, who claims that before the summer of 135 BC Hipparchus, for his first series of star observations, might have used the obliquity of 23° 55′, which points to a 12′ error, while after 135 BC, for the second series, 23° 40′, which points to a −3′ error[57].

*Afterword*

It is a great surprise that this simple regression method served with distinction as recently as in the 1920s, when Georges Lemaître (1927), Edwin Hubble (1929), and Willem de Sitter (1930) used it to find the slope of the recession velocity of the remote galaxies mapped against their distance from the Sun, now known as the Hubble constant[58]. In his famous diagram, Hubble not only depicted the data but also boldly drew a regression line or even two (see Fig. 8).

In the light of the fact that by the turn of the 20th Century the OLS regression analysis was developed to quite a high level of sophistication, mainly by the efforts of Karl Pearson and his student George Udny Yule in the 1890s[60], this approach from the side of the leading astronomers seems to be quite surprising, but also seems understandable when a researcher does not have special computational tools at his disposal.

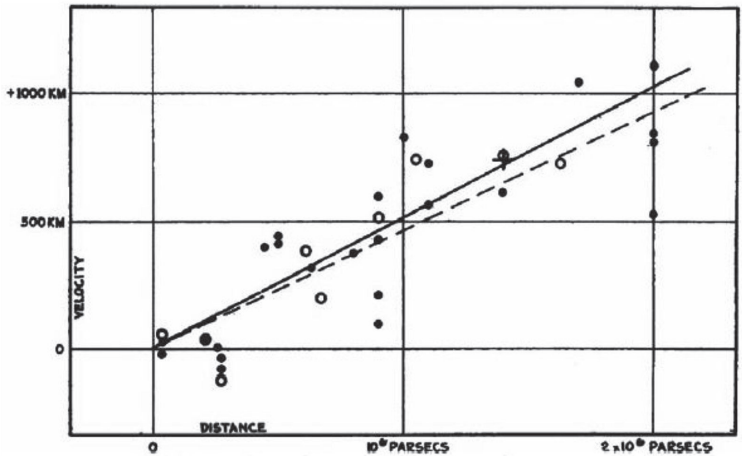

Fig. 8

The original velocity-distance relation derived by E. Hubble (1929) in two ways. The slope of the bold line, 525 km s$^{-1}$ Mpc$^{-1}$, later became known as the 'Hubble constant'. The method of computing was the simple regression analysis discovered by Newton[58].

*Summary*

In his search for the exact tropical year, Newton introduced an embryonic linear regression analysis. Not only did he perform the averaging of a set of data, 50 years before Tobias Mayer, but summing the residuals to zero he *forced* the regression line to pass through the average point. He also distinguished between two inhomogeneous sets of data and might have thought of an *optimal* solution in terms of bias, though not in terms of effectiveness. The method's simplicity led to its being applied again in the series of cosmology papers in the late 1920s. From Newton's analysis it also follows that Hipparchus' equatorial ring was misplaced by 7′, which was a matter of concern to modern science. The only visible result of his *Yahuda MS 24 D* endeavours was the tropical year that he adopted in his *Theory of the Moon's Motion* (finished on 1700 February 27, and published in 1702 by David Gregory)[61].

*Acknowledgements*

To a large extent this text is based on our earlier publication[33]. We are grateful to the following colleagues while noting their contributions: to Herbert Prinz for pointing to the letter exchange between Euler and Mayer and to Simon Cassidy for a reference to Thomas Harriot's works; to Dennis Duke (Florida University) and Robert van Gent (University of Utrecht) who made available to us the 17th-Century sources; to Todd Timberlake (Berry College) who advised on the composition of the paper; and to Joan Griffith (Annapolis MD) and Sarah Olesh (Vancouver BC) who offered suggestions on the style of this paper.